\documentclass[prb,aps,reprint,groupedaddress,showpacs]{revtex4-1}
\usepackage{amsmath}
\usepackage{graphicx}
\usepackage{epstopdf}
\usepackage{dcolumn}
\usepackage{bm}
\usepackage{cases}
\usepackage{multirow}

\begin{document}

\title{ Model of the Longitudinal Spin Seebeck Coefficient of InSb in a Magnetic Field}

\author{Nicholas A. Pike}
\email[]{Pike.55@osu.edu}
\author{David Stroud}

\affiliation{Department of Physics, The Ohio State University, Columbus, OH 49210}

\date{\today}








\begin{abstract}
 We develop a simple  theory for the longitudinal spin Seebeck effect in n-doped InSb in an external magnetic field.  We consider spin-$1/2$ electrons in the conduction band of InSb with a temperature gradient parallel to the applied magnetic field.  In the absence of spin-orbit interactions, a Boltzmann equation approach leads to a spin current parallel to the field and proportional to the temperature gradient.  The calculated longitudinal spin Seebeck coefficients oscillates as a function of magnetic field B; the peak positions are approximately periodic in 1/B.  The oscillations arise when the Fermi energy crosses the bottom of a Landau band. 
\end{abstract}

\pacs{ 71.70.Dj, 71.70.Ej, 72.20.Pa}

\maketitle

\section{Introduction}
The spin Seebeck effect refers to the generation of spin currents by an applied temperature gradient, or to the resulting voltage often induced by the so-called inverse spin Hall effect (ISHE).   The effect can be further categorized as either longitudinal or transverse.   In the longitudinal spin Seebeck effect, both the spin orientation and the spin current are parallel to the temperature gradient.   In the transverse spin Seebeck effect, a voltage difference is generated perpendicular to the temperature gradient.    A number of recent experiments have  demonstrated the occurrence of a longitudinal or a transverse spin Seebeck effect in a variety of materials~\cite{jaworski2012,uchida2010,uchida2008,bauer}. The materials involved can be metallic ferromagnet's, magnetic insulators, and even a doped non-magnetic semiconductor (Te-dopoed InSb) in a strong magnetic field~\cite{jaworski2012,uchida2010,uchida2008}.   Several papers have discussed possible explanations for such behavior but, to our knowledge, no quantitative model has been presented for n-doped InSb~\cite{bauer,takezoe2010,scharf}.

In this paper, we present a simple model calculation for spin transport in InSb in the presence of a temperature gradient and an external magnetic field.    Our model is basically a simple treatment based on the Boltzmann equation, but applied to the bands formed by the Landau levels in an n-type semiconductor when there is a strong magnetic field parallel to the temperature gradient.  The model readily leads to longitudinal spin transport.

The band structures of InSb and other zinc-blende semiconductors have been extensively investigated, both theoretically and experimentally.  Early theoretical studies by Kane~\cite{kane}, Dresselhaus~\cite{dresselhaus1955}, and Parmenter~\cite{parmenter} explain the effects of symmetry on the conduction band electronic states.  In other early studies, the effects of a magnetic field on the band structure of InSb were investigated by Roth {\it et al.}~\cite{roth1967,roth1968} and by Pidgeon {\it et al.}~\cite{pidgeon1969}. These theoretical and experimental studies led to a better understanding of the beats observed in Shubnikov-de Haas oscillations in III-V semiconductors~\cite{alsmeier}.  Other experiments showed that the lowest conduction band state in InSb has the spherically symmetric $\Gamma_6$ symmetry~\cite{dresselhaus1955,dresselhaus1955a}, and that the effective mass of the conduction band electrons is only a small fraction of the free electron mass~\cite{dresselhaus1955a}. 

  The remainder of the paper is organized as follows:   In Section II, we briefly review the relevant macroscopic transport equations describing the heat, electronic, and spin transport.    In Section III, we present a theory for these transport coefficients based on a microscopic Hamiltonian combined with the Boltzmann equation.   The Hamiltonian includes the Landau Hamiltonian for electrons in a magnetic field and the Zeeman interaction between the spins and the magnetic field.  The Boltzmann equation is then linearized, and solved to yield the  thermoelectric and longitudinal spin Seebeck coefficients.   In Section IV, we present numerical solutions of this model for the various transport coefficients as a function of magnetic field at a temperature $T=4.5K$.  In Section V, we give a brief concluding discussion.  An Appendix gives explicit expressions for the various Onsager coefficients.

\section{Macroscopic Transport Equations}
We begin by writing down the appropriate macroscopic transport equations for the system of interest, which we visualize as a doped semiconductor such as n-InSb in a magnetic field ${\bf  B}$ taken parallel to the $z$ axis.  In this case, there are three current densities to consider: the heat current density ${\bf J}_Q$, and the charge current densities ${\bf J}_+$ and ${\bf J}_-$ for spin up and spin down charge carriers.   

\begin{figure}[t]
  \centering
    \includegraphics[width=0.5\textwidth]{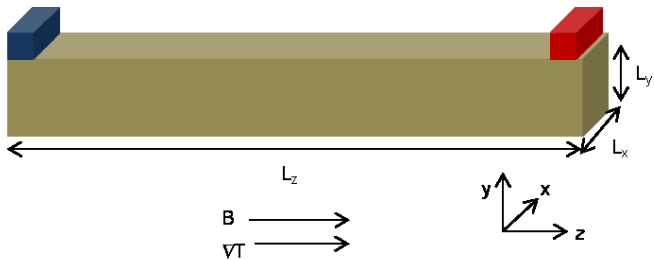}
  \caption{(Color online) Schematic of the idealized experimental configuration, showing the orientation of the magnetic field and thermal gradient, and the positions of the voltage probes used to measure the spin Seebeck coefficient in InSb.  The thermal gradient and applied magnetic field are oriented along $z$,  and the transverse voltage difference is measured between the ends of either of the two bars at the ends of the sample, which are oriented in the  $x$ direction.  The voltage drop along $z$ can also be measured by attaching leads between the two bars.}
\label{fig:theory_setup}
\end{figure}

These currents are expected to be linearly related to corresponding fields, which will be $-{\boldsymbol \nabla}T$, and $-{\boldsymbol \nabla}\mu_{\pm}$, where $T$ is the temperature and the $\mu_{\pm}$ are the electrochemical potentials for spin up and spin down charge carriers.   As noted by Brechet and Ansermet~\cite{brechet} (see also Bauer {\it et al.}~\cite{bauer} and  Uchida {\it et al.}~\cite{uchida}), the linear Onsager relations between these currents and fields may be written as
\begin{eqnarray}
\label{eq:transcoeffs1}
\left(\begin{array}{c}{\bf J}_Q \\ {\bf J}_+ \\ {\bf J}_- \end{array}\right) = \left(\begin{array}{ccc}L_{QQ} & L_{Q+} & L_{Q-} \\
    L_{+Q} & L_{++} & L_{+-} \\
    L_{-Q} &  L_{-+} & L_{--} \end{array}\right)\left(\begin{array}{c}-{\boldsymbol \nabla}T\\ -{\boldsymbol \nabla}\mu_+\\ -{\boldsymbol \nabla}\mu_-\end{array}\right)
\end{eqnarray}

Instead of the currents ${\bf J}_+$ and ${\bf J}_-$, it may be more convenient to consider the charge current density ${\bf J}_e = {\bf J}_++{\bf J}_-$  and the spin current density ${\bf J}_S = (\hbar/2q)({\bf J}_+-{\bf J}_-)$ (where $q = -e$ is the charge of the current carriers and we assume a spin $\hbar/2$ per carrier).   Similarly, rather than the $-{\boldsymbol \nabla}\mu_\pm$, it is more convenient, following Valet and Fert~\cite{valet}, to introduce the quantities $\mu_{av}$ and $\Delta\mu$ via the relation
\begin{equation} \label{eq:drivingforce}
\mu_\pm = \mu_{av} \pm \Delta\mu + qV.
\end{equation}
Here $\mu_{av}$ is the average of the two chemical potentials at zero applied voltage,  $V$ is the electrostatic potential, and $\Delta\mu = \frac{1}{2}(\mu_+-\mu_-)$.

We can now write down the linear transport equations in terms of these new fields and currents.  The result is
\begin{eqnarray}
\label{eq:transcoeffs}
\left(\begin{array}{c}{\bf J}_Q \\ {\bf J}_e \\ {\bf J}_S \end{array}\right) = \left(\begin{array}{ccc}L_{QQ} & L_{Qe} & L_{QS} \\
    L_{eQ} & L_{ee} & L_{eS} \\
    L_{SQ} &  L_{Se} & L_{SS} \end{array}\right)\left(\begin{array}{c}-{\boldsymbol \nabla}T \\ {\bf \cal E} \\ -\frac{\hbar}{2q}{\boldsymbol \nabla}(\Delta\mu) \end{array}\right),
\end{eqnarray}
where the various $L$ coefficients are all certain linear combinations of the coefficients in Eq.~\eqref{eq:transcoeffs1} and ${\cal E} = {\bf E} + \frac{{\boldsymbol \nabla}\mu}{e}$ is the effective electric field.  

Eq.~\eqref{eq:transcoeffs} applies if the spin polarization is parallel to the direction of the spin current.   If the spin polarization is not parallel to the direction of spin current flow, then there are three spin current vectors, to be called ${\bf J}_{S,i}$ with $i = x$, $y$, and $z$, corresponding to current densities of the $x$, $y$, and $z$ components of electron spin.   In this case, Eq.~\eqref{eq:transcoeffs} should be replaced by a $5\times 5$ matrix equation, corresponding to the five current densities ${\bf J}_Q$, ${\bf J}_e$, and the three ${\bf J}_{S,i}$.   Since we will not consider this situation in the present paper, we will not write down this equation explicitly.

\section{Theory for Transport Coefficients in InSb}

Next, we present a theory for some of the above transport coefficients in an n-type semiconductor, such as Te-doped InSb, in a magnetic field.   Our goal is to model experiments carried out on a sample geometry similar to that shown in Fig.~\ref{fig:theory_setup}.  In Fig.~\ref{fig:theory_setup}, we assumed that the sample is a rectangular prism having edges $L_x$, $L_y$, and $L_z$ ( $L_x$, $L_y \ll L_z$).    The top face is assumed to lie parallel to the  $xz$ plane and leads are attached to either end of the sample, as shown, so that any electrical current would flow in the $z$ direction.    A uniform magnetic field ${\bf B}$ and a uniform temperature gradient ${\boldsymbol \nabla}T$ are assumed to be applied in the $z$ direction.   

\subsection{Electronic energies and wave functions}

The conduction band of InSb is non-degenerate, and the low-lying electronic states in this band have the spectrum of a free electron (of effective mass $m^*$) in a magnetic field.   The spin-independent part of the effective-mass Hamiltonian describing these states is thus  
\begin{equation}\label{eq:elec_magfield}
H_0 = \frac{1}{2m^*}\left[+(-i\hbar\frac{\partial}{\partial x}+ q B y)^2-\hbar^2\frac{\partial^2}{\partial y^2}-\hbar^2\frac{\partial^2}{\partial z^2}\right],
\end{equation}
where we have used SI units and a gauge such that the vector potential ${\bf A} = (-By,0,0)$, where ${\bf B}={\boldsymbol \nabla}\times{\bf A}$ is the applied magnetic field.  

The solutions of the spin-independent Hamiltonian given in Eq.~\eqref{eq:elec_magfield} are standard.  The total energy is a function of a wave vector $k_z$ and the Landau level index $n$, and can be written as  
\begin{equation}\label{eq:energy}
E_{n}(k_z)= \frac{\hbar^2 k_z^2}{2 m^*}+\left( n+\frac{1}{2}\right) \hbar \omega_c,
\end{equation}
where $\omega_c = eB/m^*$ is the cyclotron frequency. Each level has a degeneracy per spin $\sigma_z$ of
\begin{equation}
\label{eq:degen} 
N_{n,\sigma_z} = \frac{L_xL_y B}{\Phi_0},
\end{equation}
 where $\Phi_0=h/2e$ is the magnetic flux quantum.

In the absence of spin-orbit interaction, the spin-dependent part of the electronic Hamiltonian, denoted $H_s$,  consists of a Zeeman interaction between the conduction electron and the applied magnetic field, which may be written as
\begin{equation}\label{eq:elec_spin_noSO}
H_s = g \mu_B {\bf B}\cdot {\boldsymbol \sigma}.
\end{equation}
Here $g$ is the electronic g-factor, which is assumed independent of the magnetic field strength, $\mu_B$ is the Bohr magneton, and ${\boldsymbol \sigma}$ is the vector of the three Pauli spin matrices for a spin-$1/2$ particle.   The eigenvalues of $H_0+H_s$ are characterized by quantum numbers $n$, $k_z$, and $\sigma_z = \pm 1/2$, and are given by
\begin{equation}\label{eq:energynoSO}
E_{n,\sigma_z}(k_z) = E_n(k_z) + g\mu_BB\sigma_z,
\end{equation}
with a spin $\sigma_z$ parallel to $z$ and a degeneracy given by Eq.~\eqref{eq:degen}.   

\subsection{Electrical, Thermal, and Spin Currents}

Next, we will obtain the various electronic transport coefficients for an n-type semiconductor such as InSb, using the Boltzmann equation.   In the presence of a magnetic field, the conduction band is broken up into many one-dimensional bands, labeled by a Landau level index $n$ and a spin index $\sigma_z$ ($\sigma_z = \pm$).  Each band is also highly degenerate, with degeneracy $N_{n,\sigma_z}$ as given in Eq.~\eqref{eq:degen}.  The Boltzmann equation for an electron of spin $\sigma_z$ in band $n$ can be written in the standard way (see, e.\ g.,  Ref.~\citenum{Ashcroft}) as 
\begin{equation}
\frac{\partial g_{n,\sigma_z}}{\partial t} + {\bf v}_{n\sigma_z}( k_z)\cdot{\boldsymbol \nabla}_{\bf r}g_{n,\sigma_z} +\frac{{\bf F}}{\hbar}\cdot{\boldsymbol \nabla}_{k_z}g_{n,\sigma_z} = \left(\frac{\partial g_{n,\sigma_z}}{\partial t}\right)_{coll}.
\label{eq:boltz1}
\end{equation}
Here $g_{n,\sigma_z}({\bf r},  k_z, t)$ is the probability that an electron in a state $ k_z$ in the $n^{th}$ band with spin $\sigma_z$ at a position ${\bf r}$ is occupied at time t, and ${\bf v}_{n\sigma_z}( k_z)$ is the velocity of an electron in the state described by $ k_z$, $\sigma_z$ and $n$.  ${\bf F}$ is the force on an electron due to an applied field.  

As is conventional, we make the relaxation time approximation so that the collision term is rewritten as
\begin{equation}
\left(\frac{\partial g_{n,\sigma_z}}{\partial t}\right)_{coll}\sim -\frac{\delta g_{n,\sigma_z}({\bf r},k_z, t)}{\tau},
\label{eq:boltz2}
\end{equation}
where $\delta g_{n,\sigma_z}$ is the deviation of $g_{n,\sigma_z}$ from its equilibrium value $g^0_{n,\sigma_z}$.  The function $g^0_{n,\sigma_z}$ is set equal to the Fermi function given by
\begin{equation}
\label{eq:fermifn}
g_{n,\sigma_z}^0(k_z) = \frac{1}{\exp[\beta(E_{n,\sigma_z}(k_z)-\mu)] +1},
\end{equation}
where $\beta = 1/k_BT$, $T$ is the temperature, $k_B$ is Boltzmann's constant, and $\mu$ is the chemical potential.  

We seek a steady state solution and thus the first term on the left hand side of Eq.~\eqref{eq:boltz1} vanishes.   We also linearize the Boltzmann equation by assuming that both ${\boldsymbol \nabla}_r g_{n,\sigma_z}$ and ${\bf F}$ are small, so that the factor $g_{n,\sigma_z}$ in both the second and the third terms of Eq.~\eqref{eq:boltz1} can be approximated as $g_{n,\sigma_z}^0$.    Combining these conditions, we obtain the linearized steady-state Boltzmann equation in the relaxation time approximation which, after simplification, is
\begin{align}
\label{eq:deltag}
-\frac{\delta g_{n,\sigma_z}(k_z)}{\tau} = & \nonumber \\
 \frac{E_{n,\sigma_z}(k_z)-\mu}{T}&\left(-\frac{\partial g_{n,\sigma_z}^0}{\partial E}\right){\boldsymbol \nabla}T\cdot {\bf v}_{n,\sigma_z}(k_z)\nonumber \\ &+  {\bf F}\cdot {\bf v}_{n,\sigma_z}(k_z)\left(\frac{\partial g_{n,\sigma_z}^0}{\partial E}\right),
\end{align}
where we evaluate $E$ at $E =E_{n,\sigma_z}(k_z)$.

We are interested in the case of an applied temperature gradient and  effective electric field oriented primary along the $z$ axis (See Eq.~\eqref{eq:transcoeffs}).   We can now express the electric, heat, and spin currents in terms of $\delta g_{n,\sigma_z}(k_z)$, as obtained from Eq.~\eqref{eq:deltag}.  Two of these expressions are given, for a spherical band, by, e.\ g., Ref.~\citenum{Ashcroft}.  These general expressions need to be modified to take account of the degeneracy of the Landau bands as we do below.  The mathematical form of the spin current density, ${\bf J}_{S},$ is similar to that of the electrical current density, ${\bf J}_e$.  

To be explicit, we can write out the current densities in the system as follows: 
\begin{align}
\label{eq:current}
{\bf J}_e &= \frac{L_z}{V} \sum_{n,\sigma_z} \int \frac{d k_z}{2\pi}(-e) N_{n,\sigma_z} {\bf v}_{n,\sigma_z}(k_z) \delta g_{n,\sigma_z}(k_z),  \\
{\bf J}_Q &= \frac{L_z}{V} \sum_{n,\sigma_z} \int \frac{d k_z}{2 \pi}N_{n,\sigma_z}\Delta E_{n,\sigma_z} {\bf v}_{n,\sigma_z}(k_z) \delta g_{n,\sigma_z}(k_z), \\
{\bf J}_{S,i} &= \frac{L_z}{V} \sum_{n,\sigma_z} \int \frac{d k_z}{2 \pi}\mu_B \langle\sigma_{n,i}\rangle N_{n,\sigma_z} {\bf v}_{n,\sigma_z}(k_z) \delta g_{n,\sigma_z}(k_z),
\label{eq:current_end}
\end{align}
where we take the integral over $k_z$ from $\pm \infty$ and $\Delta E_{n,\sigma_z} = E_{n,\sigma_z}(k_z)-\mu$.  

In Eqs.~\eqref{eq:current} -~\eqref{eq:current_end}, ${\bf J}_Q$ is the heat current, ${\bf v}_{n,\sigma}(k_z)=\hbar^{-1}{\boldsymbol \nabla}_{k}E_{n,\sigma_z}(k_z)$ is the velocity of the electron in the band labeled by $(n,\sigma_z)$, and $\langle\sigma_{n,i}\rangle$ is the expectation value of the $i^{th}$ component of spin in the band $(n,\sigma_z)$ ($i = x,y,z$).   In the absence of spin-orbit interaction, only ${\bf J}_{S,z}$, that is, the current density associated with the $z$ component of spin, is non-zero.  

For an electron in the conduction band the velocity, ${\bf v}_{n,\sigma_z}(k_z) = v_{n,\sigma_z}(k_z){ \hat{z}}$ is given by
\begin{align}\label{vspin}
v_{n,\sigma_z}(k_z) &= \frac{\hbar k_z}{m^*}  \nonumber \\
&=\pm \frac{\hbar}{m^*} \sqrt{\frac{2m^*}{\hbar^2}(E_{n,\sigma_z}(k_z)-E^0_{n,\sigma_z})},
\end{align}
where the $+$ and $-$ signs apply when $k_z>0$ and $k_z< 0$, respectively, and $E_{n,\sigma_z}^0$ is defined as the minimum energy for the band (n, $\sigma_z$)  given by
\begin{equation}
\label{eq:nsig0}
E_{n,\sigma_z}^0 = \left(n+\frac{1}{2}\right)\hbar\omega_c +\sigma_z g\mu_BB.
\end{equation}
 
In order to calculate the various transport coefficients in Eqs.~\eqref{eq:current} -~\eqref{eq:current_end}, we need the chemical potential $\mu$.    $\mu$ can be calculated given the conduction electron density $\rho=N_e/V$, where $N_e$ is the total number of conduction electrons in volume $V$.    For the present case, the chemical potential $\mu$ is obtained from 
\begin{equation}\label{eq:fermi_energy_int}
\rho =\frac{B}{\pi \Phi_0}\sum_{n,\sigma_z} \int_0^\infty g^0_{n,\sigma_z}(k_z)dk_z
\end{equation}
where Eq.~\eqref{eq:fermi_energy_int} is an implicit equation for $\mu(T,B)$. 

Since the experiments of Ref.~\citenum{jaworski2012} are done at a very low temperature ($T = 4.5K$), we have approximated $\mu$ (or equivalently, the Fermi energy $E_F$) by its value at $T = 0$.  In this case, $g_{n,\sigma_z}^0$
is just a step function, and $E_F$ is given implicitly by
\begin{equation}\label{eq:fermi_energy}
\rho= \sum_{n,\sigma_z} \frac{B}{\pi \Phi_0}\left(\frac{2 m^*}{\hbar^2}\left(\mu -E_{n,\sigma_z}^0\right)\right)^{\frac{1}{2}},
\end{equation}
where $E_{n,\sigma_z}^0$ is defined in Eq.~\eqref{eq:nsig0} and the sum runs only over Landau bands with nonzero electron occupation. 

\begin{figure}[t]
  \centering
    \includegraphics[width=0.5\textwidth]{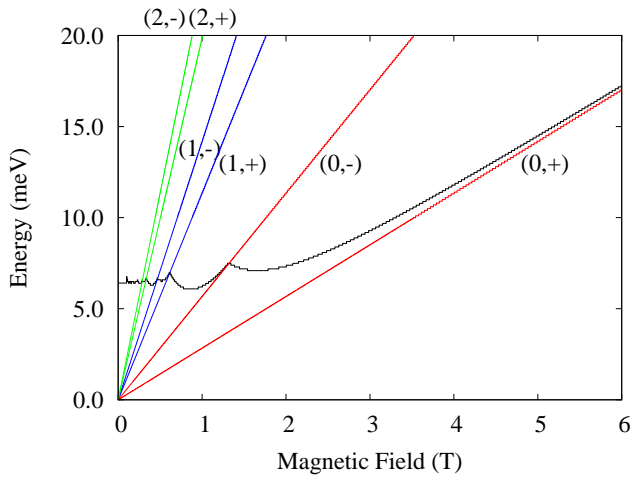}
  \caption{(Color Online) Calculated Fermi energy at $T = 0K$ plotted as a function of an applied magnetic field along the $z$ direction as indicated in Fig.~\ref{fig:theory_setup}.  We also show the first three Landau bands ($n= 0,1,2$); each band is labeled by its band index $n$ and spin $\sigma_z$ as (n, $\sigma_z$), where $\sigma_z = \pm$.
Straight lines correspond to the minima of the various Landau sub-bands, as labeled in the Figure.  Scalloped curve represents the T= 0 Fermi energy $E_F$ as a function of
magnetic field.  In this figure, $E_F$ is calculated neglecting spin-orbit coupling.}
\label{fig:fermi_energy_bands}
\end{figure}

\begin{figure}[t]
        \centering
                \includegraphics[width=0.5\textwidth]{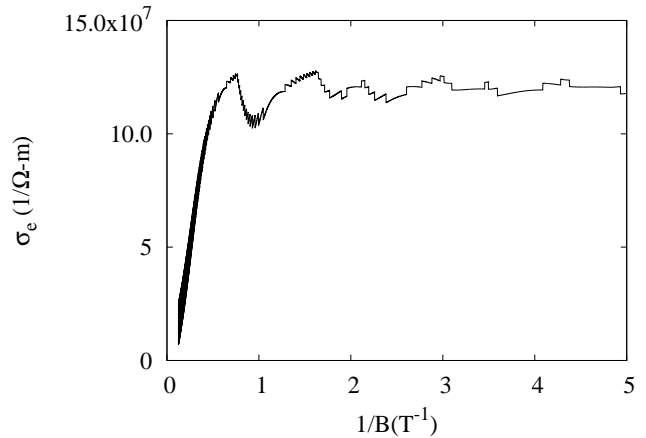}
               \caption{The calculated electrical conductivity $\sigma_{e}$, given in Eq.~\eqref{app:econ}, plotted as a function of the inverse magnetic field $1/B$ at $T=4.5 K$, and neglecting spin-orbit interactions.  The maxima in the conductivity occur when the Fermi energy crosses the bottom of the Landau band as plotted in Fig.~\ref{fig:fermi_energy_bands}.  The highest-field maximum occurs at approximately $B= 1.2T$, which corresponds to the Fermi energy crossing the $(0,-)$ Landau level.  This curve is calculated neglecting spin-orbit interaction and using the $T = 0$ Fermi energy.}
                \label{fig:conductivity}
\end{figure}

\subsection{Transport Coefficients}

We obtain the transport coefficients of interest by using Onsager's linear relationship between the currents and the forces generating the currents~\cite{onsager}.  For the present problem, this relation is given by Eq.~\eqref{eq:transcoeffs}. These may be written in condensed form as
\begin{equation}
\label{eq:onsager}
J_{i} = \sum_j L_{ij} F_j,
\end{equation}
where $i = e, Q, S$ runs over the three currents in the system and $j$ runs over the forces  acting on the conduction electrons.  In this paper, we consider only longitudinal spin currents, i.\ e., we assume that only $\langle \sigma_{z}\rangle \neq 0$.   The explicit form of the measured transport coefficients can then be obtained by combining Eq.~\eqref{eq:transcoeffs} with Eqs.~\eqref{eq:deltag} -~\eqref{vspin}.

For example, the electrical conductivity, $\sigma_e$, is given by the Onsager coefficient $L_{ee}$ (see Fig. \ref{fig:conductivity}).   Similarly, the thermal conductivity, $\kappa$, is given by~\cite{Ashcroft}
\begin{equation}\label{eq:thermalcon}
\kappa =\frac{L_{QQ} L_{ee}-L_{Qe}L_{eQ}}{L_{ee}}
\end{equation}
where the Onsager coefficients are given in Appendix~\ref{app1} and shown in Fig.~\ref{fig:thermalcon}.
\begin{figure}[t]
        \centering
               \includegraphics[width=0.5\textwidth]{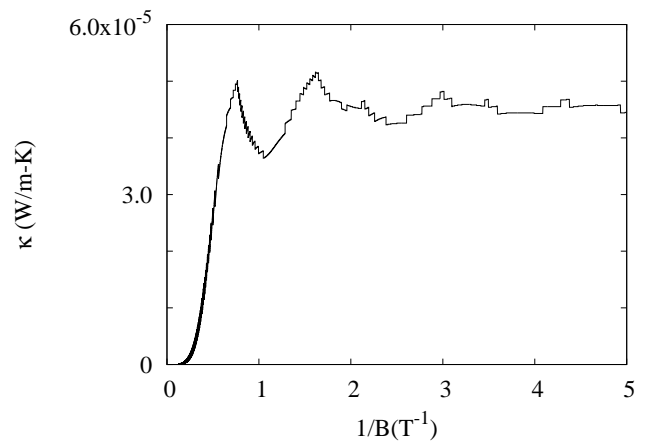}
               \caption{Calculated thermal conductivity, $\kappa$, given in Eq.~\eqref{eq:thermalcon}, plotted as a function of $1/B$ at $T = 4.5 K$.  The maxima in $\kappa$ occur, like those of $\sigma_e$, when the field-dependent Fermi energy crosses the bottom of a Landau level. This curve is calculated assuming no spin-orbit interaction and the values of the $T=0K$ Fermi energy. }  
                \label{fig:thermalcon}
\end{figure}

 The thermopower $\alpha$ is generally defined as the ratio of the $z$ component of the electric field to the negative of the thermal gradient (also assumed to be in the $z$ direction) under the condition of zero electrical current in the $z$ direction.   We write this condition as  ${\cal E} = \alpha(-{\boldsymbol \nabla}T)_{{\bf J}_e=0}$~\cite{Ashcroft}. It is readily shown that the $\alpha$ can be expressed in terms of the Onsager coefficients as
\begin{equation}\label{eq:thermopower}
\alpha = -\frac{L_{eQ}}{L_{ee}}.
\end{equation}

We can also calculate coefficients related to spin transport driven by a temperature gradient.   For the case of longitudinal spin transport, we need to calculate the coefficient $L_{SQ}$, as defined in Eq.~\eqref{eq:transcoeffs}.   $L_{SQ}$ is the ratio of the longitudinal spin current density to the applied temperature gradient, i.\ e.,  $J_{S,z} = -L_{SQ}\nabla_zT$ under conditions such that all other currents and forces are negligible.   In Fig.~\ref{fig:alphazz}, we have plotted $L_{SQ}$ as a function of inverse magnetic field.

\begin{table}[t]
\begin{center}
\begin{tabular}{| l |   l| c | }
\multicolumn{3}{ c }{Assumed values for the physical properties of InSb} \\
\hline
Quantity           &   Value  &  Ref. \\                   
\hline
 $g$                     & $-49.0$           & ~\citenum{madelung} \\
\hline
$m^*$               & $ 0.013 m_e$    &~\citenum{dresselhaus1955a}\\
\hline
$\tau$                   &  $\approx1\times10^{-7}\ s  $& ~\citenum{Whalen1969,Grisar1976} \\
\hline 
\end{tabular}
\caption{Numerical parameters used in the calculation of the Onsager coefficients given in Appendix~\ref{app1}.  
 The estimate $\tau \sim 10^{-7}$ s is typical of that found in n-doped InSb samples at $T = 4.5 K$~\cite{Whalen1969,Grisar1976}.}
\label{tab1}
\end{center}
\end{table}

\section{Numerical Results}

We now turn to numerical results based on the present simplified model.    We first calculate the chemical potential $\mu(T,B)$ at $T = 0K$, assuming parameters appropriate to the conduction band of InSb and the experiments of Ref.~\citenum{jaworski2012}, as given in Table~\ref{tab1}.   

The resulting Fermi energy is  shown in Fig.~\ref{fig:fermi_energy_bands} as function of $B$, assuming a conduction electron density of $\rho=3.7 \times 10^{15}$ cm$^{-3}$, as used in the experiments of Ref.~\citenum{jaworski2012}.  The results show, as already obtained in Ref.~\citenum{jaworski2012}, that $\mu(T=0, B)$ is a non-monotonic function of $B$, with discontinuous changes in slope wherever the minimum of one of the spin sub-bands rises through the Fermi energy and becomes unoccupied.  

Given the Fermi energy, or at finite temperature the chemical potential $\mu(T,B)$, we can calculate a variety of transport coefficients. Here we calculate the components of electrical and thermal conductivities, and of the thermopower,  parallel to the field, under the appropriate experimental conditions as described above.  We also calculate the transport coefficient $L_{SQ}$ [Eq.~\eqref{eq:transcoeffs}], which represents the spin current density in the $z$ direction per unit applied temperature gradient in the $z$ direction. Expressions for the relevant Onsager coefficients are given in Appendix~\ref{app1}.  All the coefficients are functions of both the applied magnetic field $B$ and the temperature $T$.   The integrals in the transport coefficients are all dominated by energies within $ k_BT$ of $E_F$, since the energy derivative of the Fermi function, which is a factor in each of the integrals, is strongly peaked near $E_F$.  

\begin{figure}[t]
        \centering
                \includegraphics[width=0.5\textwidth]{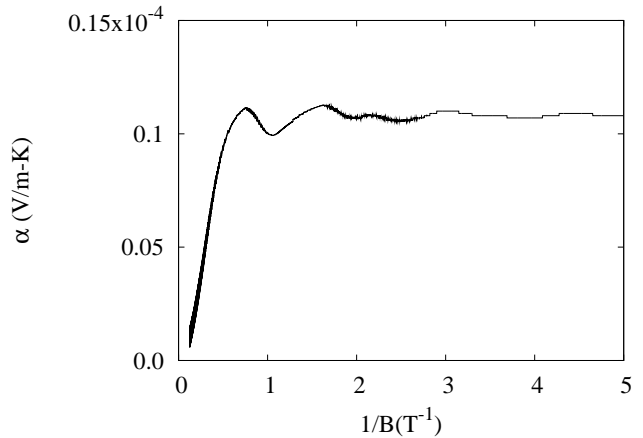}
                \caption{ The calculated thermopower, as obtained using Eq.~\eqref{eq:thermopower}, plotted versus $1/B$ at $T = 4.5 K$.   The maxima occur in the same manner as in Figs.~\ref{fig:conductivity} and~\ref{fig:thermalcon}. This curve is calculated assuming no spin-orbit interaction and the values of the $T=0K$ Fermi energy.}
                \label{fig:thermopower}
\end{figure}

\begin{figure}[t]
        \centering
               \includegraphics[width=0.5\textwidth]{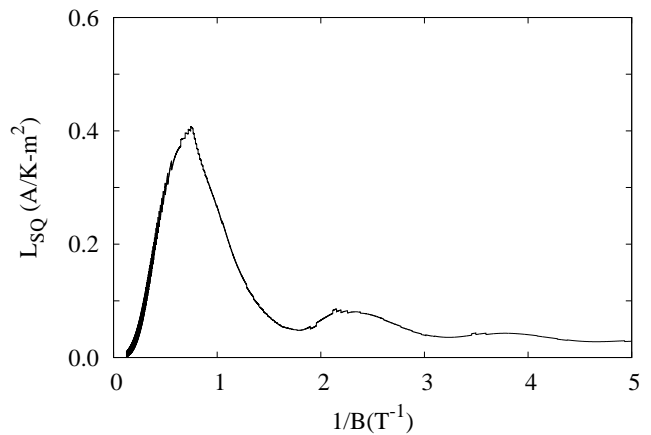}
                \caption{ The calculated longitudinal spin Seebeck coefficient $L_{SQ}$ given by Eq.~\eqref{app:lsq}, plotted versus $1/B$ at $T = 4.5 K$ but using the $T=0K$ Fermi energy.  The maxima occur when the minima of the various Landau sub-bands cross the Fermi energy, as in Figs.~\ref{fig:conductivity} -~\ref{fig:thermopower}.}
                \label{fig:alphazz}
\end{figure}

The results of these calculations are shown in Figs.~\ref{fig:conductivity} -~\ref{fig:alphazz}.  In each case, we have plotted the transport coefficients at $T= 4.5K$  as functions of the inverse magnetic field.   The various numerical parameters used in the calculations are given in Table~\ref{tab1}.  We plot the transport coefficients in this manner in order to show  that the positions of the peaks in these quantities vary periodically with $1/B$. The oscillations are related to the de Haas- van Alphen oscillations normally seen in the magneto-transport coefficients of metals~\cite{shoenberg}.

\section{Discussion and Conclusions}

In this paper, we have given a simple model for the longitudinal spin Seebeck coefficient in InSb.   In our model, the electronic energy levels of n-type InSb in a magnetic field are given as Landau levels and the various electronic transport coefficients, including the longitudinal spin Seebeck coefficient, are obtained from a simple Boltzmann equation approach for each Landau subband.  The oscillations of this coefficient in a magnetic field occur when the Fermi energy crosses the minima of the various Landau subbands as the magnetic field is varied.    

While our model holds, in principle, for any temperature $T$,  we have carried out the calculations of the transport coefficients only at low $T$ ($T \sim 4.5$ K) and specifically calculated the chemical potential at $ T = 0K$ (See Fig.~\ref{fig:fermi_energy_bands}).  While the difference between the chemical potential at $T=0$ and $T=4.5K$ is small, it could affect both the magnitude and position of the Landau level crossings.  This change could give quantitatively different results, but he qualitatively picture of the oscillations would remain the same. 

Finally, we discuss how our simple model might be modified to produce a transverse spin Seebeck effect.     The present model omits spin-orbit interaction, which is known to have a large effect on the band structure of InSb and similar compound semiconductors.  The spin-orbit interaction couples the spatial momentum to various components of the electronic spin. In particular, some forms of this interaction couple momenta in one direction with spin components in other directions.   Such coupling could lead to expectation values of the spin vector which are tilted relative to the electronic momentum.   This could, in turn, produce a nonzero value of ${\bf J}_{S,x}$ and ${\bf J}_{S,y}$ along the $z$ direction.  If the spin vector is tilted relative to the direction of spin current, this will lead to a transverse electric field via the inverse spin Hall effect (ISHE)~\cite{saitoh}.   An ISHE electric field would also be produced if the spin is oriented in the $z$ direction but the corresponding spin current has a component in the $x$ or $y$ direction.   In a future paper, we plan to present a model for this transverse spin Seebeck effect based on this picture.

\section{Acknowledgments}
 This work was supported by the Center for Emerging Materials at The Ohio State University, an NSF MRSEC (Grant No.\ DMR0820414).  The authors would like to thank  C.\ Jaworski, J.\  P.\  Heremans, E.\ Johnston-Halperin, and R.\ C.\ Myers for valuable discussions.

\appendix
\section{Onsager Coefficients}
\label{app1}
Here we give expressions for the various Onsager coefficients discussed and calculated in the text.  In our model, the Onsager coefficients are
\begin{align}
L_{ee} &= \frac{L_z}{V} \sum_{n,\sigma_z}\int_{-\infty}^{\infty} \frac{d k_z}{2 \pi} e^2N_{n,\sigma_z} \times  \label{app:econ}\\ \nonumber
 & \qquad  [v_{n,\sigma_z}(k_z)]^2 \tau \frac{\partial g^0_{n,\sigma_z}(E,T)}{\partial E}; \\
L_{eQ} &=  \frac{L_z}{V} \sum_{n,\sigma_z}\int_{-\infty}^{\infty} \frac{d k_z}{2 \pi} e N_{n,\sigma_z}[v_{n,\sigma_z}(k_z)]^2 \times \\ \nonumber &\tau \frac{\Delta E}{T}\frac{\partial g^0_{n,\sigma_z}(E,T)}{\partial E};\\
L_{Se,i} &= -\frac{L_z}{V} \sum_{n,\sigma_z}\int_{-\infty}^{\infty} \frac{d k_z}{2 \pi} e \mu_B \langle \sigma_{i}(k_z) \rangle N_{n,\sigma_z} \times \\ \nonumber &[v_{n,\sigma_z}(k_z)]^2\tau \frac{\partial g^0_{n,\sigma_z}(E,T)}{\partial E};\\
L_{SQ,i} &= -\frac{L_z}{V} \sum_{n,\sigma_z}\int_{-\infty}^{\infty} \frac{d k_z}{2 \pi} \mu_B\langle\sigma_{i}(k_z)\rangle N_{n,\sigma_z}\times \label{app:lsq}\\ \nonumber &[v_{n,\sigma_z}(k_z)]^2\tau  \frac{\Delta E}{T} \frac{\partial g^0_{n,\sigma_z}}{\partial E};\\ 
L_{QQ} &= \frac{L_z}{V} \sum_{n,\sigma_z}\int_{-\infty}^{\infty} \frac{d k_z}{2 \pi} N_{n,\sigma_z}\frac{(\Delta E)^2}{T}\times \label{app:lqq}\\ \nonumber&[v_{n,\sigma_z}(k_z)]^2 \tau  \frac{\partial g^0_{n,\sigma_z}}{\partial E}, \nonumber
\end{align}
where $L_{eQ} = L_{Qe}$ and $\Delta E = E_{n,\sigma_z}(k_z) - \mu$.   In Eqs.~\eqref{app:econ} -~\eqref{app:lqq}, the derivative $\partial g_{n,\sigma_z}^0(E,T)/\partial E = -\beta e^{\beta(E-\mu)}/[e^{\beta(E-\mu)} + 1]^2$, with $\beta = 1/(k_BT)$, and  $E = E_{n,\sigma_z}(k_z)$.  During numerical calculations the integrals given above are converted to integrals over energy using  the relationship $dk_z  = dE/[dE/dk_z]$.   All the integrals are dominated by the energy range within $k_B T$ of the Fermi energy, because $\partial g^0_{n,\sigma_z}/\partial E$ is strongly peaked around $E = E_F$.   In practice, the integrands all become vanishingly small beyond an energy of $\sim 3k_BT$ on either side of $E_F$.  In all the above expressions, $v_{n,\sigma_z}(k_z)$ is obtained from Eq.~\eqref{vspin}.

\end{document}